# Exploring the ethical sensitivity of Ph.D. students in robotics

Linda Battistuzzi[1], Lucrezia Grassi [1] and Antonio Sgorbissa [1]

[1] University of Genova, Genova, Italy
`linda.battistuzzi@unige.it`

**Abstract.** Ethical sensitivity, generally defined as a person's ability to recognize ethical issues and attribute importance to them, is considered to be a crucial competency in the life of professionals and academics and an essential prerequisite to successfully meeting ethical challenges. A concept that first emerged in moral psychology almost 40 years ago, ethical sensitivity has been widely studied in healthcare, business, and other domains. Conversely, it appears to have received little to no attention within the robotics community, even though choices in the design and deployment of robots are likely to have wide-ranging, profound ethical impacts on society. Due to the negative repercussions that a lack of ethical sensitivity can have in these contexts, promoting the development of ethical sensitivity among roboticists is imperative, and endeavoring to train this competency becomes a critical undertaking. Therefore, as a first step in this direction and within the context of a broader effort aimed at developing an online interactive ethics training module for roboticists, we conducted a qualitative exploration of the ethical sensitivity of a sample of Ph.D. students in robotics using case vignettes that exemplified ethical tensions in disaster robotics.

## 1 Background

Ethical sensitivity (also termed ethical awareness or moral sensitivity/sensibility), a concept that first emerged in moral psychology almost four decades ago [1], is generally defined as a person's ability to recognize ethical issues when they emerge in practice, while at the same time attributing importance to those same issues [2, 3]. Ethical sensitivity thus involves an intuitive, reflexive ability to recognize the ethically relevant facets of situations, but also deliberate attention to such aspects [2, 4]. It is considered a key competency in the life of professionals and academics and an essential prerequisite for meeting ethical challenges.

Moral functioning, however, requires additional competencies [1, 5], and ethical sensitivity is neither obvious nor stable. Indeed, research suggests that people largely differ in their responsiveness to the ethical dimension of daily life [4, 6, 7]. While some individuals display a marked ability to recognize and respond to ethical issues and violations, others are more likely to be unaware or "blind" to such challenges [7]. This variation has been attributed to factors including personal experience and social-ization, suggesting that ethical sensitivity can be promoted and shaped by learning [7].

The impact of ethical blindness can be both extensive and profound. Research on ethically blind individuals has found that they can inadvertently behave in ways that are not consistent with their own values and principles [8]. Unless people have the ability to identify the ethical facets of everyday life and understand why they matter, no ethical issues will exist for them. [2]. If the ethical issues at stake are not identified, there will be no need to engage with ethical problem-solving [9]. Thus, ethical sensitivity is generally viewed as the first stage in ethical decision-making and ethical behavior.

In light of this, ethical sensitivity has been extensively studied in a number of professional and academic domains including, nursing, medicine, and business [2, 3, 10, 11]. Conversely, it appears to have received little to no attention within the robotics community, even though the choices that roboticists make when designing and deploying robots are likely to have wide-ranging, deep ethical impacts on society. As the IEEE's Technical Committee on Robot Ethics warns, robotics research and technology prompt urgent questions of ethical importance [12]. In the field of care robotics, for instance, scholars have expressed concerns about patient autonomy [13] and informed consent, quality of data management, human contact, infantilization and equality of access [14, 15] . In disaster robotics, ethical concerns include questions related to privacy, responsibility, safety and trust [16].

Due to the negative repercussions that a lack of ethical sensitivity can have in these and other contexts in which robots are likely to play increasingly prominent roles, promoting the development of ethical sensitivity among roboticists is imperative, and endeavoring to train this competency becomes a critical undertaking. Therefore, as a first step in this direction, we conducted a qualitative exploration of the ethical sensitivity of a sample of Ph.D. students in robotics using case vignettes that exemplified ethical questions in disaster robotics.

## 2    Methods

Within the context of a wider effort aimed at developing an interactive, online ethics training module for Ph.D. students, 29 students enrolled in the first year of the Ph.D. program at the Dept. of Bioengineering, Robotics and Systems Engineering at the University of Genoa were invited to participate in our study. Eight students declined; the reason for their refusal was not investigated.

After providing their informed consent, the remaining 21 participants were asked to fill out a brief sociodemographic survey (including information about their age, academic qualifications, and previous training in ethics). Then they read through an interactive online story and reacted in writing (in English or Italian, as they preferred) to four short vignettes contained within the story. The online story, which was written in English, had already been piloted for clarity. Usability testing had also been conducted and reported [17].

A qualitative approach was chosen as previous studies have shown that qualitative studies with vignettes on ethical factors and dilemmas can elicit understandings that are not captured in quantitative paradigms [18]. Hence, vignettes have the potential to generate insightful findings in this context [19].

This study was approved by the Ethics Committee of the University of Genoa (Comitato Etico d'Ateneo Decision n. 2021/46).

## 2.1 The story and the vignettes

The main character in the story is a drone expert working at a prominent university who has received a grant from a private company to develop a new kind of multicopters. At the beginning of the story, we learn that the character has joined rescue operations at the site of a devastating earthquake. The story, which we described in detail elsewhere [17] revolves around a series of ethically problematic situations that arise before and during the operations. (Before the story begins, an introductory section of the activity provides students with a brief explanation of what is meant by "ethics" and "ethical issue".). Some of those situations become decision points to which the students' attention is explicitly called. Depending on the choices that students make when faced with those ethical challenges, the story takes different directions, leading to different paths and different endpoints. Each path requires that the students make a minimum of four, and as many as eight, decisions.

In other parts of the story, students are presented with vignettes describing ethical issues that the main character in the story is faced with. The students are then asked to react to the situations described, stating whether they perceive any actual or potential ethical concern and, if they do, to explain in writing what the problem is.

In qualitative research such as this, vignettes are "concrete examples of people and their behaviors on which participants can offer comment or opinion" [20]. The issues presented by the vignettes in our study are derived from four different papers on ethical issues in rescue robotics: our scoping review [16], Murphy and Moats' study outlining a model for responsible robotics innovation for disaster response, based on evidence from 36 case studies [21], Harbers' and colleagues study on the ethical landscape of robot-assisted search and rescue [22], and van Wynsberghe and Comes's paper on drones in humanitarian contexts and robot ethics [23].

The length of vignettes is also known to affect the quality and quantity of data elicited, as longer texts can generate careless responses owing to loss of interest by participants, while shorter ones have been consistently found to elicit useful responses [19] Our vignettes were therefore kept brief.

Because the story follows different paths depending on the choices students make as they progress through it, 21 students encountered the first vignette, 13 the second, five the third, and 21 the fourth.

The first vignette (reported in **Table 1** below) is presented at the very beginning of the story and revolves around the ethical tension between the business priorities of private research sponsors, research integrity and independence, and questions related to how safe and beneficial novel technologies must be before they are deployed in the real world.

**Table 1.** Vignette 1

*You are a robotics engineer at the Dept. of Robotics and Artificial Intelligence of a highly regarded European university. Most of your team's work focuses on disaster robots. A few years ago, you were awarded significant funding by a private company. The company commissioned you to develop an innovative multicopter with aquatic landing capability and interchangeable aerial and marine camera views. You and some of the more senior members of your lab have obtained your Remote Pilot Certificates and regularly train with the multicopters. Over time you have met several fire department and civil protection officials and have joined in their training exercises. In order to propel your team's efforts forward, as your funders often remind you, it is now crucial to move from high-fidelity simulations to actual rescue missions. Although your system still has some usability and battery issues, you need to find opportunities to show the responder community how valuable your multicopters can be, especially for shallow searches and working over water.*

The second vignette (shown in **Table 2** below) builds on the previous one, emphasizing the question of experience with a novel technology, which is closely related to the problem of premature deployment. It also introduces greater complexity, asking students to think more specifically about uncertainty and risk.

**Table 2.** Vignette 2

*You find the Incident Commander at IC headquarters and tell him about the possibility of aquatic landings as he is walking towards his truck. You explain what this could mean in terms of supporting rescue operations in submerged areas, and he quickly grasps the significance of what you are saying. He asks you whether this could in some way increase the possibility of failures and you have to acknowledge that your experience with water landings is limited.*

The third vignette, reported in **Table 3**, raises issues of reliability and trust.

**Table 3.** Vignette 3

*As you finish setting up your gear, you realize that the extra generators you thought you had brought along are not in the truck. You make a mental note to check who oversaw backup material but are relieved that the additional batteries and components to complete any repairs are all there; the problem may not be so serious. Also, should the need arise, you could probably ask the firefighter drone team to lend you one.*

And finally, the fourth vignette (see **Table 4**) introduces the question of privacy as well as issues of self-promotion, transparency, trust and community engagement.

**Table 4.** Vignette 4

*The Incident Commander leaves, asking you to continue operations and use your multicopters to take pictures and film of specific areas. Most survivors who are able to walk on their own have been evacuated but casualties could be incidentally filmed or photographed, nonetheless. Taking your cue from the Incident Commander's request, you remind your team that civilian privacy must be preserved. Then you decide to look for the Public Information Officer (PIO) to see whether a press release can be organized to inform the community and brief the media about your drones and what they are being used for.*

## 2.2 Analysis of participant responses

Thematic analysis, a widely used method to analyze qualitative data [24], was used to identify patterns of meaning in the participants' responses.

## 3 Results

### 3.1 Sociodemographic features of participants

Among the 21 participating students, females were a minority (8/21). Most (17/21) fell in the 25-34 age bracket, and almost all (19/21) had a Master of Science degree as their highest academic qualification. Finally, 3/21 had attended an ethics course or had had some form of ethics training previously (see **Table 5** below).

**Table 5** Sociodemographic features of participants

| Participant ID | Age | Gender | Highest qualification | Previous ethics training |
|---|---|---|---|---|
| 5 | 25-34 | Female | Master of Science | No |
| 7 | 35-44 | Female | Ph.D. | No |
| 9 | 18-24 | Female | Master of Science | No |
| 11 | 25-34 | Male | Master of Science | No |
| 13 | 25-34 | Male | Master of Science | No |
| 17 | 25-34 | Male | Ph.D. | No |
| 19 | 25-34 | Male | Master of Science | Yes |
| 23 | 25-34 | Female | Master of Science | No |
| 25 | 25-34 | Male | Master of Science | No |
| 27 | 25-34 | Male | Master of Science | No |
| 29 | 25-34 | Female | Master of Science | No |
| 31 | 25-34 | Female | Master of Science | No |
| 33 | 25-34 | Male | Master of Science | No |
| 35 | 18-24 | Male | Master of Science | No |
| 35T | 25-34 | Female | Master of Science | Yes |
| 37 | 25-34 | Female | Master of Science | Yes |
| 41 | 25-34 | Male | Master of Science | No |
| 45 | 25-34 | Male | Master of Science | No |

| 47 | 25-34 | Male | Master of Science | Yes |
| 48T | 25-34 | Male | Master of Science | No |
| 51 | 25-34 | Male | Master of Science | No |

### 3.2 Participant responses

#### Vignette 1

Regarding Vignette 1, 20/21 participants responded that the situation presented was ethically problematic, one stated that they saw no ethical problem. All of those who recognized the ethical dimension of the vignette reported concerns related to the risks, safety, and harm associated with the novel drone technology. One also wondered whether the university team had enough experience to deploy the drones safely in actual rescue settings as opposed to a simulation:

"*Using the system in an actual rescue scenario is risky and potentially dangerous, considering that there are still some issues. In addition, the lab members who will join the rescue mission lack experience and knowledge compared to firemen and civil protection officials*". (ID5)

While acknowledging that deploying an immature technology is ethically problematic, another student added that simulations are not sufficient to test a novel technology:

"*Using a technology that is not ready and still has some unresolved issues is ethically wrong, as it may cause more harms than benefits. On the other hand, some issues may be impossible to solve only using simulations, but there may be the need to test the technology "on site", with the conditions and variables that only reality can present*". (ID31)

Four participants also recognized that research funding from a private sponsor may be ethically problematic if it puts pressure on academics to deploy a novel technology prematurely. Three of these participants emphasized the risks that private funding can pose in terms of research independence and integrity:

"*Private companies can fund research projects in order to influence the direction of research.*" (ID33)

"*The development of the multicopter could be negatively influenced by the financial interests of the company.*" (ID37)

"*Pressure from private sponsors could be an ethical problem, they could use the funding to control when and how the technology is released.*" (ID45)

One participant acknowledged that there is a balance that needs to be found, as private funding can play a crucial role in supporting research:

"*There is an ethical concern regarding research funding. While on the one hand the professional rescuers and university researchers try to develop new rescue tools for the common good, the private company is looking for a return on its investment. On the other hand, funding from the company is what has made the research project possible in the first place.*" (ID19)

### Vignette 2

Moving on to Vignette 2, 9/13 students found the situation presented of ethical concern, while four did not. Among the latter four, two simply stated that they saw no ethical problem, while two elaborated:

"*No, the Incident Commander has been informed that there may be increased failures, he has all the pros and cons and can decide whether to proceed with the new multicopters on water or not.*" (ID13)

"*The engineer has to clearly explain the harms and benefits associated with using the multicopters on water and provide estimates of the chance of success. If he thinks using these drones could create problems for the rescue operations, he should be honest with the Incident Commander, who will then consider the pros and cons and make a decision.*" (ID37)

Among the students who instead recognized the ethical complexity of the situation described in the vignette, one emphasized how using a new technology when its usefulness is yet to be proven can, among other things, waste precious time:

"*It's as if I was asking the Incident Commander to test my technology during the rescue operations. The time needed could be used more usefully to focus on important priorities in that situation.*" (ID25)

Another stated that providing the Incident Commander with the most accurate and reliable information possible has significant ethical weight:

"*As the person responsible for the design of the multicopters and the one who has most experience using them, I have a moral obligation to provide the most accurate information possible to support the Incident Commander's decision. The information must be honest and verifiable.*" (ID19).

Three students clearly understood that the situation involved questions of risk and decision-making under uncertainty; after premising that there is a crucial lack of information, however, each went in a different direction. The first concluded that the Incident Commander cannot make a good decision given the lack of information:

*"The situation is ethically problematic because the technology is immature and hasn't been tested enough under the conditions described. This puts pressure on the Incident Commander who doesn't have enough information to make a good decision."* (ID31)

The second took a step back, pointing out how the final responsibility for making a decision rests with the Incident Commander:

*"Yes, you are aware that there is a risk of failures, and you don't know whether deploying the multicopters on water is worth the additional time, effort and logistic strain. At the end of the day, however, the decision is the Incident Commander's.*" (ID51)

The third, instead, went one step further, balancing the risk and the potential for a positive outcome:

*"The problem here is uncertainty, there isn't enough information to make a decision, but if using an unproven technology can improve the chances of saving lives, I would implement it."* (ID48)

### Vignette 3

Coming to Vignette 3, two of the five students who encountered it reported that they viewed the situation as ethically concerning:

*"If I take another team's extra generator because I forgot mine, and then they need it, I have put them in a difficult situation because of my carelessness."* (ID25)

*"You are allowed to be at the disaster site because you can help rescue operations, not to interfere with them. Thinking of counting on the rescuers' resources instead of your own team's shows that you are not reliable.*" (ID37)

Among the other three students, one simply stated that there was no issue, the other two suggested that it is important to be straightforward about which resources are actually available so that they can be managed optimally:

*"I think it's important to coordinate with the other rescuers to decide priorities and how to use resources together.*" (ID41)

*"In order to avoid any kind of problem, I would be very clear about my equipment and let the Incident Commander know about what I have available so he can decide how my team should participate in rescue operations."* (ID45)

**Vignette 4**

Finally, regarding Vignette 4, 19/21 students recognized that there was an ethical dimension in the situation described, while two did not. Among those 19, ten reiterated what was already stated in the vignette, i.e., that civilian privacy must be preserved. One added that the data recorded should only be available to rescuers for rescue purposes:

"*The privacy of citizens must always be preserved, and rescuers should only collect images to help in the rescue operations. None of the images should be shared outside the rescue effort*" (ID37)

Another student pointed out that it is not just the privacy of victims and civilians that is at risk:

"*There may be privacy issues for both rescuers and victims.*" (ID23)

One also noted how privacy issues can have mental health consequences for victims and their families both in the short and in the long term:

"*The risk here is sharing information with people who are not involved in rescue operations would violate victims' privacy. This could have consequences on their personal life later on. What if a video were posted on social media? The people affected could see it, even after a year or more, reopening wounds and causing psychological harm.*" (ID51)

Several students (8/19) also commented on the character's decision to get in touch with the Public Information Officer to organize a press conference. Most of these students (7/8) took a negative view of this decision, considering it a form of advertising or self-promotion:

"*By informing the community about our drones and the work we are doing, we are exploiting a tragic event to promote the company that sponsors us and our project.*" (ID45)

According to these students, the press conference also distracts the main character from rescue operations and is a waste of precious time:

"*It is more important to focus on the rescue operations rather than show our multicopters to the media.*" (ID5).

"*Is this the right moment to advertise my team's work? I would put that off until the rescue operations are over.*" (ID7).

*"I think that the ethical issue here is twofold. On one hand, we are wasting precious time thinking about the media instead of focusing on the instructions the Incident Commander has given us. On the other hand, we are taking advantage of a disaster situation, to advertise our work with the media."* (ID47)

One student viewed the press release as a way to keep the community informed:

*"The press release is correctly issued in order to inform the community."* (ID13)

## 4 Discussion

To explore their ethical sensitivity, we asked a sample of Ph.D. students at the Dept. of Bioengineering, Robotics, and Systems Engineering at the University of Genoa to react to four vignettes describing ethical issues that may occur when robots are deployed in disaster scenarios. This was conducted within the context of a more extensive effort aimed at developing an interactive, online ethics training module.

The first vignette exemplifies situations in which roboticists look for opportunities to promote their technology in response settings. As Murphy and Moats point out, roboticists are often motivated by a genuine desire to support rescue operations by sharing their knowledge and expertise [21]. However, if the technology is immature or will not clearly benefit the rescue effort in terms of bringing value to the responders' goals, priorities, and activities, deploying it may be unethical [21].

Another issue that emerges in the first vignette is that of pressure from corporate funders. Private funding seldom comes with no strings attached. Indeed, corporate and industry funders often choose to support research precisely because they expect to influence the direction that research will take [25]. Maintaining academic integrity and independence can require conducting a risk-benefit analysis, considering the alignment of corporate sponsors with the recipient's mission and values. Researchers also need to be aware of situations in which their commitment to advancing research can be jeopardized by industry objectives [25].

While almost all our study participants were able to recognize the ethical problems associated with the deployment of an immature technology in terms of safety and potential harms, only 4/20 also commented on the issue of corporate pressure. Just one of those four mentioned the need to strike a balance between the benefits of private funding and the risks it may involve. These findings suggest that the PhD students in our sample had limited awareness of the ethical complexities involved in dealing with private sector funding.

The second vignette builds on the previous one. While in the first vignette the main character acknowledges that the multicopters "still have usability and battery issues", here the problem is that the actor also "has limited experience with landings on water". The situation is complicated by the fact that this information is directly shared with the Incident Commander, who has the ultimate responsibility of making a decision.

According to Murphy and Moats, emergency managers like the Incident Commander in our vignette generally evaluate a robot based on three factors: availability, suitability, and risk. Responders' tolerance for performance and operational risks will depend on the specific application. For instance, they may have a greater tolerance for a drone that might fail when flying through a building in areas that cannot be reached in any other way, than for a ground robot that could seriously jeopardize all rescue operations if it were to malfunction [21]. More generally, the riskier the situation, the more likely that stringent risk standards are ethically justifiable [26].

Among our participants, it was interesting to notice how some saw no ethical quandary at all in this vignette, as long as the Incident Commander was provided with all the information available. Only three clearly understood that pushing a technology too soon, without sufficient testing is risky, and it forces the Incident Commander to make a decision without sufficient information, which is also ethically problematic. Each of the students reacted differently to this: one concluded that the Incident Commander could not possibly make a good decision given the uncertainty of the situation; the second seemed less concerned with the Incident Commander's position, perhaps trusting in the Commander's competence; the third instead seemed to be more tolerant of risk as long as saving lives was a possible outcome. The degree of variation in the responses to this vignette sits well with research suggesting that tolerance to risk and uncertainty varies significantly between individuals and between cultures, so it is likely that diverging views will emerge in discussions about the ethics of risks [26]

The issues in the third vignette have to do with reliability and trust. As Murphy and Moats remind us, whenever robots are deployed in rescue efforts, those robots will be expected to be available throughout the emergency. Therefore, in order to provide a positive contribution to rescue operations and establish a relationship of trust with professional rescuers, roboticists should be confident that their robot can function reliably and that sufficient resources are at hand to keep it in operation for as long as necessary [21].

Only 2/5 of the students who encountered this vignette directly or indirectly mentioned reliability in their comments; none discussed trust. Yet trust - the belief that another will follow through on your behalf - is a key construct when we think about teamwork and technology use [27]. In teams, trust among teammates often predicts various outcomes, including team performance and job satisfaction [28]. Trust toward a given technology has also been found to be an important predictor of use with that technology [29]. Successfully integrating robots into the workflow of rescue efforts requires that rescue robots prove trustworthy but also that roboticists establish themselves as reliable partners on the field.

In the fourth vignette, the prominent ethical question is privacy. The goals and capabilities of rescue robots determine increased information gathering, both purposeful and incidental. This may be personal information about rescue workers, such as images or data about their physical and mental stress levels, but also about victims or people living or working in the disaster area [30]. Thus, although increased information flow is widely accepted as appropriate for emergencies and disasters, information flows across robots generate more opportunities for privacy infringements [16]. Ro-boticists should be familiar with the relevant implicit or explicit expectations or policies. In some jurisdictions, for instance, responders expect all imagery to be handled with a chain of custody to preserve civilian privacy [21].

Privacy was recognized as ethically relevant by almost all of the students in their comments. Most were clear about the risks that sharing images of the disaster site could entail for victims and other civilians, although, interestingly, only one mentioned that rescuers' privacy can also be infringed during operations.

Roughly half of the students remarked that the character's decision to organize a press conference was tantamount to advertising or self-promotion. Only one considered that the local community as well as the public at large ought to be informed that drones were being deployed for rescue purposes. Drones flying overhead can be perceived as invasive, and it may be difficult if not impossible to understand who is flying them or for what purpose, which may cause concern among those in the area [23]Transparent communication is essential to ensure that the technology is used in ways that are not only beneficial but are also perceived as such. Communication is also important to prevent or at least curtail misinformation on the part of the media, which can mislead public expectations of what robots can and cannot do, ultimately hampering trust in the technology itself [16]

There are often methodological questions about the validity of vignettes, and one could reasonably debate how much the situation depicted in a given vignette truly represents the phenomenon under study [31]. However, we believe that grounding our vignettes in research that specifically focuses on the questions at hand has helped us ensure that there is a clear link between those vignettes and the experiences being explored. In addition, asking students whether our vignettes involved an ethical issue, using the term "ethical" in our wording, may have emphasized the possibility that the situation did indeed contain an ethical dimension, which the students might have not considered had the question been phrased differently [31]. We acknowledge that this makes our approach susceptible to social desirability, although for all four vignettes at least one student (and sometimes several more) reported that they recognized no ethical concern.

Overall, our findings suggest that the Ph.D. students who participated in the study were mostly able to identify ethical concerns linked to safety and privacy issues that may arise when robots are deployed in rescue settings. They were far less aware, however, when it came to issues of reliability, trust, transparency and handling pressures from corporate funders. These issues have significant impact well beyond the specific domain of rescue robotics. Although caution should always be exercised when drawing conclusions from qualitative explorations, these results emphasize the importance of providing ethics training to young roboticists if we are to encourage their ability to engage with ethical challenges in their professional lives.

## Acknowledgments

We are grateful to the students whose generous participation made this study possible.

This work was supported by the Italian Ministry of Education, University and Research through the DIONISO project (Research code: B82C12000070001; National Civil Service Code SCN_00320), under the Smart Cities and Communities and Social Innovation program.

## References

1. Rest JR (1986) Moral development: Advances in research and theory. Praeger, New York
2. Ineichen C, Christen M, Tanner C (2017) Measuring value sensitivity in medicine. BMC Medical Ethics 18:
3. Jordan J (2007) Taking the first step toward a moral action: a review of moral sensitivity measurement across domains. J Genet Psychol 168:323–359
4. Jordan J (2009) A social cognition framework for examining moral awareness in managers and academics. J Bus Ethics 84:237–258
5. Tanner C, Christen M (2014) Moral Intelligence—A framework for understanding moral competences. In: Christen M, van Schaik C, Fischer J, et al (eds) Empirically informed ethics: Morality between facts and norms. Springer International Publishing AG, pp 119–136
6. Reynolds S (2008) Moral attentiveness: Who pays attention to the moral aspects of life? The Journal of Applied Psychology 93:1027–1041
7. Schmocker D, Tanner C, Katsarov J, Christen M (2019) An Advanced Measure of Moral Sensitivity in Business. European Journal of Psychological Assessment 36:1–10
8. Bazerman M, Tenbrunsel A (2011) Blind spots: Why we fail to do what's right and what to do about it., 1st Ed. Princeton University Press
9. Clarkeburn H (2002) A test for ethical sensitivity in science. Journal of Moral Education 439–453
10. Miller J, Rodgers Z, Bingham J (2014) Moral Awareness. In: Agle B, Hart D, Thompson J, Hendricks H (eds) Research companion to ethical behavior in organizations. Edward Elgar Publishing, Cheltenham, UK, pp 1–43
11. Kim Y, Park J, You M, et al (2005) Sensitivity to ethical issues confronted by Korean hospital staff nurses. Nursing Ethics 12:595–605
12. IEEE IEEE Technical Committee for Robot Ethics - Scope and Objectives. https://www.ieee-ras.org/robot-ethics. Accessed 2 Aug 2022
13. Sparrow R (2016) Robots in aged care: A dystopian future? AI & Society 31:445–454
14. Sharkey A, Sharkey N (2010) Granny and the robots: Ethical issues in robot care for the elderly. Ethics and Information Technology 14:27–40. https://doi.org/10.1007/s10676-010-9234-6
15. Sparrow R, Sparrow L (2006) In the hands of machines? The future of aged care. Minds and Machines 16:141–161

16. Battistuzzi L, Recchiuto CT, Sgorbissa A (2021) Ethical concerns in rescue robotics: a scoping review. Ethics and Information Technology. https://doi.org/10.1007/s10676-021-09603-0

17. Battistuzzi L, Grassi L, Recchiuto C, Sgorbissa A (2021) Design and usability testing of a text-based simulation for ethics training in disaster robotics. In: Building and Evaluating Ethical Robotic Systems - IROS Workshop

18. Braun V, Clarke V (2013) Successful Qualitative Research: A Practical Guide for Beginners. Sage, London

19. Azman H, Mahadir M (2017) Application of the Vignette Technique in a Qualitative Paradigm. GEMA Online Journal of Language Studies 17:

20. Hazel N (1995) Elicitation techniques with young people. Social Research Update Winter:

21. Murphy R, Moats J (2021) Responsible Robotics Innovation for Disaster Response. In: Building and Evaluating Ethical Robotic Systems - IROS Workshop

22. Harbers M, de Greeff J, Kruijff-Korbayová I, et al (2017) Exploring the Ethical Landscape of Robot-Assisted Search and Rescue. Springer, Cham, pp 93–107

23. van Wynsberghe A, Comes T (2020) Drones in humanitarian contexts, robot ethics, and the human–robot interaction. Ethics in Information Technology 22:43–45. https://doi.org/10.1007/s10676-019-09514-1

24. Braun V, Clarke V (2006) Using thematic analysis in psychology. Qualitative research in psychology 3:77–101

25. Fabbri A, Lai A, Grundy Q, Bero LA (2018) The influence of industry sponsorship on the research agenda: A scoping review. American Journal of Public Health 108:. https://doi.org/10.2105/AJPH.2018.304677

26. Green B (2019) Six Approaches to Making Ethical Decisions in Cases of Uncertainty and Risk. In: Markkula Center for Applied Ethics. https://www.scu.edu/ethics/focus-areas/technology-ethics/resources/six-approaches-to-making-ethical-decisions-in-cases-of-uncertainty-and-risk/. Accessed 3 Aug 2022

27. You S (2017) Technology with embodied physical actions: Understanding interactions and effectiveness gains in teams working with robots. In: Conference on Human Factors in Computing Systems - Proceedings

28. de Jong BA, Elfring T (2010) How does trust affect the performance of ongoing teams? the mediating role of reflexivity, monitoring, and effort. Academy of Management Journal 53:. https://doi.org/10.5465/amj.2010.51468649

29. Wu K, Zhao Y, Zhu Q, et al (2011) A meta-analysis of the impact of trust on technology acceptance model: Investigation of moderating influence of subject and context type. International Journal of Information Management 31:. https://doi.org/10.1016/j.ijinfomgt.2011.03.004

30. Harbers M, de Greeff J, Kruijff-Korbayová I, et al (2017) Exploring the Ethical Landscape of Robot-Assisted Search and Rescue. pp 93–107

31. Wilks T (2004) The Use of Vignettes in Qualitative Research into Social Work Values. Qualitative Social Work 3:. https://doi.org/10.1177/1473325004041133